\documentclass[]{article}
\usepackage{graphicx,amsmath}
\usepackage[hmargin=3cm,vmargin=3cm,bindingoffset=0.5cm]{geometry}
\usepackage{amsmath,amssymb,dsfont}
\usepackage{authblk}
\usepackage{url}
\usepackage{xcolor}
\usepackage[normalem]{ulem}
\title{Spatio-temporal Patterns of Indian Monsoon Rainfall}
\author[1]{Adway Mitra}
\author[2]{Amit Apte}
\author[2]{Rama Govindarajan}
\author[2]{Vishal Vasan}
\author[3,4]{Sreekar Vadlamani}
\affil[1]{School of Electrical Sciences, Indian Institute of Technology, Bhubaneswar, India}
\affil[2]{International Center for Theoretical Sciences (ICTS), Bangalore, India}
\affil[3]{Center for Applicable Mathematics, Tata Institute of Fundamental Research (TIFR-CAM), Bangalore, India}
\affil[4]{Department of Statistics, Lund University, Sweden}

\begin{document}

\maketitle

\begin{abstract}

The primary objective of this paper is to analyze a set of
canonical spatial patterns that approximate the daily rainfall across the Indian region, as identified in the companion paper where
we developed a discrete representation of the Indian summer monsoon rainfall using state variables with spatio-temporal coherence maintained using a Markov Random Field prior. In particular, we use these spatio-temporal patterns to study the variation of rainfall during the monsoon season. Firstly, the ten patterns are divided into three families of patterns distinguished by their total rainfall amount and geographic spread. These families are then used to establish `active' and `break' spells of the Indian monsoon at the all-India level. Subsequently, we characterize the behavior of these patterns in time by estimating probabilities of transition from one pattern to another across days in a season. Patterns tend to be `sticky': the self-transition is the most common. We also identify most commonly occurring sequences of patterns. This leads to a simple seasonal evolution model for the summer monsoon rainfall. The discrete representation introduced in the companion paper also identifies typical temporal rainfall patterns for individual locations. This enables us to determine wet and dry spells at local and regional scales. Lastly, we specify sets of locations that tend to have such spells simultaneously, and thus come up with a new regionalization of the landmass.

\end{abstract}

%%%%%%%%%%%%%%%%%%%%%%%%%%%%%%%%%%%%%%%%%%%%%%%%%%%%%
%%%%%%%%%%%%%%%%%%%%%%%%%%%%%%%%%%%%%%%%%%%%%%%%%%%%%
\section{Introduction}
\label{sec:intro}

The complex dynamical nature of Indian summer monsoon is not only of interest to a large section of researchers, but also has a profound impact on one of the world's largest economies. Understanding the spatio--temporal variations of the monsoon is an important, mathematically challenging problem. In the companion paper
\cite{mitra2018monsoon1}, we established a theoretical model for the rainfall by introducing some latent variables which encoded local rainfall conditions and spatio--temporal patterns. By setting appropriate parameter values, we also obtained a fixed number of spatial patterns of rainfall, and we extracted prominent patterns which were used to {\it approximate} the daily spatial distribution of rainfall. Such an analysis integrating small and large scale variations is a novel approach towards capturing extreme variations of rainfall patterns over the Indian landmass, ranging from scanty to abundant rainfall. Additionally, the obtained patterns exhibit better coherence, and are readily interpretable. 

To the best of our knowledge, there have been very few studies related to identifying spatial patterns of weather (and in particular rainfall), and their evolution across a season. One such work is~\cite{weathertypes} where six \emph{weather types}, each specified by a spatial pattern of daily low-altitude horizontal winds, are identified using the K-means clustering technique over the Pacific Ocean. The spatial patterns are considered to be snapshots of intra-seasonal or inter-seasonal oscillations. Properties of the different weather types, such as their distribution across seasons, and the associated characteristics of rainfall and cloud-cover are also studied. Regarding spatial patterns of monsoon rainfall over India, the work~\cite{miso} computes two Empirical Orthogonal Functions (EOFs) and the corresponding principal components of daily spatial rainfall distribution every day in a season. 

Having established the spatial patterns, it is natural to study them to extract relevant information. In this paper, we observe that the spatial patterns can further be grouped into families using certain homogeneous features. For instance, we observe that a few spatial patterns clearly correspond to the pre-onset and post-retreat period of monsoon, thus combining them into a group provides a clearer picture. Additionally, an important aspect of any statistical study is the test for robustness of the estimates outside the training/estimation dataset. In doing so, we observe an interesting relationship between the Hamming similarity index, and the intensity of rainfall.

A major theme of this paper is the time evolution of spatial patterns over a monsoon season, which may help capture the pan-Indian progression of rainfall. In particular, we study the transition relations between these patterns which may allow one to predict the spatial distribution of rainfall a few days into the future based on the current spatial distribution. The trends we observe are extensively consistent with the observations of climate scientists~\cite{gadgilvariability}. In particular, we are able to provide a mathematical framework for many {\it known} features of the monsoon progression. For instance, we observe that the spatial patterns tend to be `sticky', i.e., the self transitions occur with high probability when compared to other transitions. 

In addition to the spatial patterns, we also study \emph{temporal} patterns and clustering of the spatial locations. Though the temporal patterns are not as readily interpretable as their spatial counterparts, they still encode a lot of information, and are later used to define local wet and dry spells.

One of the features of Indian monsoon, which has been the focus of many research reports is the presence of ``active'' and ``break'' spells (see ~\cite{josephspells,rajeevanspells,singhspells,krishnamurthyspells,goswamispells}). These spells have been attributed to various physical variables like the cloud cover, pressure and wind pattern and wind intensity (\cite{ramamurthybreak,goswamispells}). However, the largely accepted way to define active and break spells in a nonparametric way is to compare the all India spatial aggregate rainfall against unit standard deviation interval around the mean of the climatological data (see \cite{krishnamurthyspells}). A similar analysis with the all Indian aggregate replaced by aggregate over the monsoon zone was suggested by \cite{josephspells,rajeevandataset}.

Recently~\cite{windspells} considered active and break spells with respect to both rainfall and wind over south-western part of peninsula.

Though the definition of spells in these papers vary, they all define these spells at the all-India scale. However, even at local or regional scales, the daily distribution of rainfall is not uniform, but tends to follow alternating dry and wet spells. Also, neighboring locations are likely to have such spells simultaneously. In this work we identify regions - each region being a set of neighboring grid-locations - where these spells tend to occur simultaneously. Furthermore, in addition to the regional wet-dry spells, we also study various characteristics of local wet and dry spells.

The rest of this paper is organised as follows: in Section \ref{sec:method}, we briefly discuss our model along with the notations and variables. Thereafter, in \ref{sec:results} we present a detailed analysis of the spatial patterns including the distribution of prominent patterns over the four monsoon months in Section \ref{sec:prominent-sp-patterns}. The focus of Section \ref{subsec:robustness} is to test the robustness of the our spatial patterns over a larger dataset. Thereafter, the time evolution of these spatial patterns constitutes the central theme of Section \ref{subsec:time-evolve}. In Section \ref{sec:actbrk} we discuss various notions of active--break spells from pan-Indian to local scales by establishing a link from the spatio--temporal patterns to active (wet) and break (dry) spells. Finally, in Section \ref{sec:discuss}, we summarize the contribution of this paper, and discuss possible future directions as natural extensions of this work.

%%%%%%%%%%%%%%%%%%%%%%%%%%%%%%%%%%%%%%%%%%%%%%%%%%%%%
%%%%%%%%%%%%%%%%%%%%%%%%%%%%%%%%%%%%%%%%%%%%%%%%%%%%%

\section{Methodology}
\label{sec:method}

Using a Markov random field model we, in \cite{mitra2018monsoon1}, developed a data-driven discrete model of the rainfall at $S$ locations, over a period of $D$ days. In this section, we shall briefly overview coarse details of this model, necessary to follow the discussion hereinafter. For details, the reader is referred to the companion paper~\cite{mitra2018monsoon1}.

We begin with the rainfall data\footnote{We have adopted the standard notation of expressing random variables with upper case characters, and their realization with corresponding lower case characters.} denoted by $x(s,t)$ for locations $s=1,\ldots,S$, and days $t=1,\ldots,D$. Additionally, we introduce the following variables:
\begin{itemize}
\item a binary variable $Z$, for each $(s,t)$, encoding the intensity of rainfall.
\item discrete clustering variables $U$ and $V$, for each day and location, encoding the spatial and temporal patterns in $Z$, and the rainfall data $x$, respectively. These are to be interpreted as tags/indices assigned to each day and location. 
\end{itemize}

Using the assignment of the index variables $V$ and $U$, we then partition the set of all spatial locations and days into spatial and temporal clusters, respectively. The total number of different tags determines the number of clusters formed. Subsequently, we pick one element from each cluster as a representative of that cluster. Such elemments will be labelled as {\it patterns}.

\begin{figure}
	\centering
	\includegraphics[width=10cm, height=6cm]{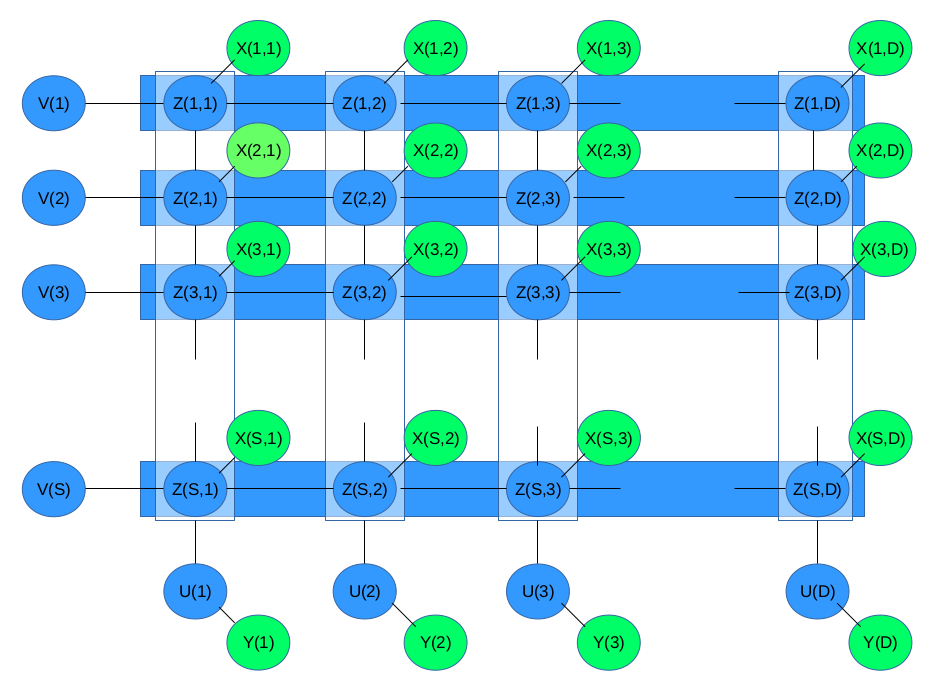}
	\caption{The proposed Graphical Model for Indian Rainfall.  Each column represents one day, each row represents a location. $Z$: latent state variable, $X$: observed rainfall volume. Horizontal edges are ``temporal", vertical edges ``spatial",  ``data edges" connect $Z$, $X$ nodes. The $U$-nodes below represent daily cluster variables, connected to all local state variables on same day denoted by the vertical rectangles. The $V$-nodes on the left represent spatial cluster variables, connected to all local state variables on same day denoted by the vertical rectangles.}
\label{fig:mrfmodel}\end{figure}

The dependence structure between these variables is elaborated via the graphical model in Figure \ref{fig:mrfmodel}. Using this structure and setting $Y(t) = \sum_{s=1}^S X(s,t)$, the joint probability density of $(Z,U,V,X)$ can be expressed as
\begin{eqnarray}
  p(Z,U,V,X) &\propto& p_u(U) \times p_v(V) \nonumber \\
&\times& \prod_{s,t}^{S,T}\prod_{t'=t-1}^{t+1}\psi_T(Z(s,t),Z(s,t'))\times \prod_{s,t}^{S,T}\prod_{s'\in \Omega(s)}\psi_S(Z(s,t),Z(s',t)) \nonumber \\
  & \times& \prod_{s,t}^{S,T}\psi_{ST}(Z(s,t),V(s))\times\prod_{s,t}^{S,T}\psi_{SS}(Z(s,t),U(t)) \nonumber \\
&\times& \prod_{s,t}^{S,T}\psi_{DZ}(Z(s,t),X(s,t))\times\prod_{t}^{T}\psi_{DU}(U(t),Y(t))
\label{eq:mrfmodel} \end{eqnarray}
The first two lines in the above equation represent the prior distribution of $U$, $V$, \& $Z$, given by $p_u$, $p_v$, and the edge potentials $\psi_T$, \& $\psi_S$ describing the joint distribution of $Z(s,t)$ via the temporal and spatial dependence. The last two lines account for the relationship between the variables $Z$, $U$, $V$ and $X$, via the edge potentials $\psi_{ST}$, $\psi_{SS}$, $\psi_{DZ}$ and $\psi_{DU}$, over the edges appearing in the graphical representation in Figure \ref{fig:mrfmodel}. For complete details about the functional form of these edge potentials, we refer the reader to the accompanying article \cite{mitra2018monsoon1}.

We note that the spatio--temporal patterns we analyze here are obtained as a function of the latent variables $(U,V,Z)$ given the data $x$.  Therefore, the central inference step involves sampling from the conditional distribution $p(Z,U,V|X=x)$, and generating spatio--temporal patterns using the estimates thus computed.

We use the standard Gibbs sampler\footnote{See \cite{mitra2018monsoon1} for details, and discussion, related to the sampling step.} to sample from the the posterior distribution $p(Z,U,V|X=x)$, wherein we begin with an initial assignment of variables $Z$, $U$ and $V$, and at each subsequent step update the variables corresponding to each node conditioned on all the other variables. The Markovian nature of our model ensures that the updation step involves conditioning only on the {\it neighboring sites} which are identified by the presence of edges. Finally, mode of the conditional distribution of $(Z,U,V)$ given $\{X=x\}$ at any stage of updation is considered to be representative of the conditional distribution $p(Z,U,V|X=x)$.

At any stage of the Gibbs sampling algorithm, we could use the $Z$, $U$ and $V$ assignments to define spatial and temporal patterns. First, we set $x(\cdot,t)$ and $Z(\cdot,t)$ for the $S$ dimensional vector representing the $x$ and $Z$ values at all the $S$ locations on day $t$, then for each spatial cluster index $u$, we define the associated spatial patterns corresponding to the rain fall data $x$ ({\it canonical rainfall pattern} (CRP)), and latent variable $Z$ ({\it canonical discretized pattern} (CDP)) as:
\begin{eqnarray}
\phi_u &=& \textrm{mean}_t\left(x(\cdot,t) : U(t) = u\right) \,, \nonumber \\
\phi_u^d &=& \textrm{mode}_t\left(Z(\cdot,t) : U(t) = u\right) \,,
\label{eq:phiu} \end{eqnarray}
where the mean/mode is taken over the days that belong to the cluster $u$. Demonstrably, since each $X(t)$ and $Z(t)$ is an $S$-dimensional vector, so are $\phi_u$ and $\phi^d_u$.
We note here that the number of spatial clusters, and thus the patterns, created by the model depends on the data and model parameters mentioned in Table 2 of \cite{mitra2018monsoon1}.

Similarly, for each temporal cluster index $v$, the associated patterns for the data $x$, called the {\it canonical time series} (CTS), and for the latent variable $Z$, called the {\it canonical discretized series} (CDS), are defined as
\begin{eqnarray}
\theta_v &=& \textrm{mean}_s\left(x(s,\cdot) : V(s) = v\right) \,, \nonumber \\
\theta_v^d &=& \textrm{mode}_s\left(Z(s,\cdot) : V(s) = v\right) \,.
\label{eq:thetav} \end{eqnarray}
where the mean/mode is taken over the locations that belong to the cluster $v$.

%%%%%%%%%%%%%%%%%%%%%%%%%%%%%%%%%%%%%%%%%%%%%%%%%%%%%
%%%%%%%%%%%%%%%%%%%%%%%%%%%%%%%%%%%%%%%%%%%%%%%%%%%%%

\section{Results: Spatial Patterns}
\label{sec:results}
Having defined our model, and the spatial patterns thus obtained, we  now present our results. The dataset used for this work was published by Indian Institute of Tropical Meteorology. It provides daily rainfall data for the period 1901-2011 at $1^{\circ}-1^{\circ}$ spatial resolution~\cite{rajeevandataset} all over India, and it is available on request.

We fit our model on daily rainfall data over $357$ locations, for $8$ years- $2000-2007$ during the $4$ months (June-September)-$122$ days. So in our study, $S=357$ and $D=122\times 8=976$.

\subsection{Prominent Spatial Patterns}\label{sec:prominent-sp-patterns}
\begin{figure}[t!]
	\centering
	\includegraphics[width=\textwidth]{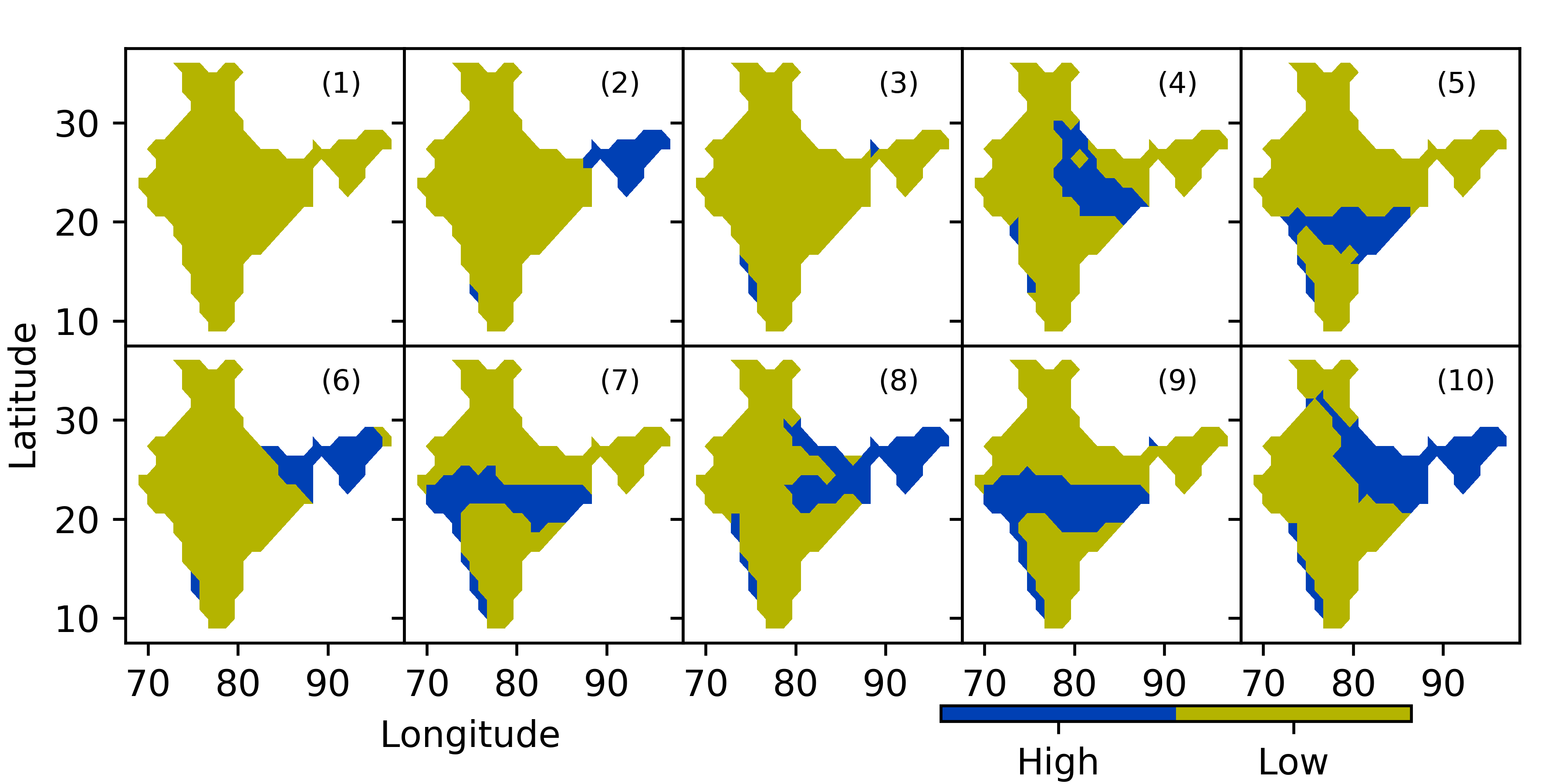}
	\includegraphics[width=\textwidth]{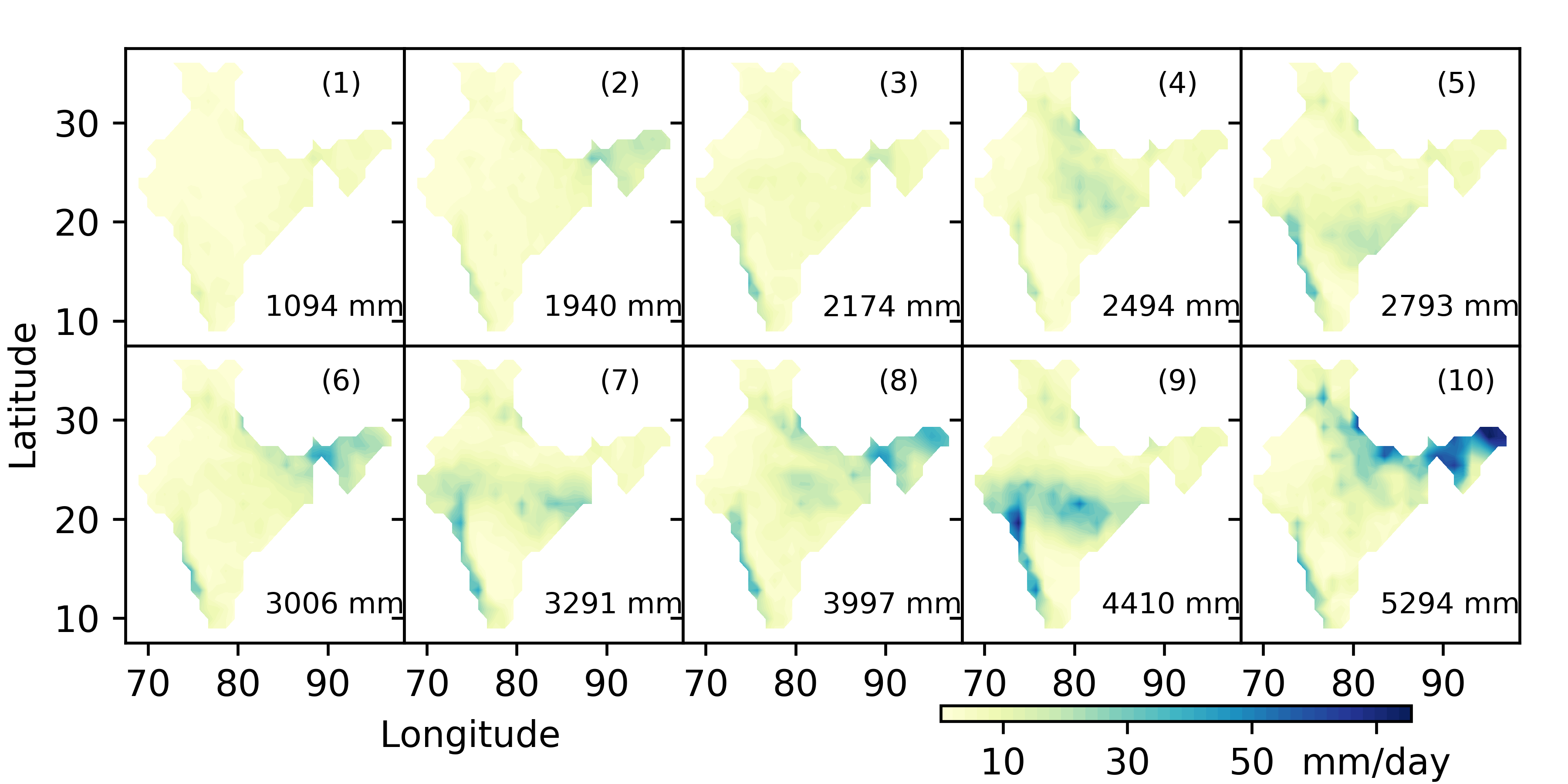}
	\caption{Prominent canonical discrete patterns (CDP) and the corresponding canonical rainfall patterns (CRP) identified by the proposed model. The numbers on bottom right show the total rainfall in mm/day for that pattern.}\label{fig:mrf-cdp-crp} 
\end{figure}

As discussed in the companion paper, the $U$-variable identifies daily clusters and their associated canonical discrete patterns of rainfall. Each cluster is associated with a canonical discretized pattern. The proposed model produces about $10$ prominent clusters, along with $10$ prominent canonical discretized patterns, $\phi_u^d$. Each of these patterns is \emph{prominent}, i.e. appears in at least $5$ of the $8$ years considered. These patterns can be shown on a map, as done in Figure~\ref{fig:mrf-cdp-crp}. 

The clustering of the days can be used to identify real-valued canonical rainfall patterns, $\phi_u$, which are analogous to $\phi_u^d$. In the lower part of Figure~\ref{fig:mrf-cdp-crp} we show the canonical rainfall patterns computed this way from the prominent clusters identified by the proposed model. These are reasonably correlated with the binary patterns shown in the upper part of Figure~\ref{fig:mrf-cdp-crp}. For each canonical rain pattern $\phi_u$, we can compute the spatially aggregated rainfall, and can  order the patterns in increasing order of aggregated rainfall.

It is not difficult to hypothesize that the first three patterns of Figure~\ref{fig:mrf-cdp-crp} are associated with the pre-onset (early June) or post-retreat (late September) period or break spells where there is no or very little rainfall except in the western coast and northeastern hilly regions. In patterns $5$, $7$, and $9$ we see rainfall over Central India (the monsoon zone) and the Western coast. Although the spatial distribution of rainfall is very similar in these three patterns, the aggregate rainfall volumes are different, which can also be seen in the corresponding $\phi_u$ patterns. These patterns are associated with the ``active spells" of Indian monsoon when rainfall is high over the Monsoon zone. In contrast, the patterns $4$, $6$, $8$ and $10$ exhibit rainfall concentrated in the Gangetic Plains of the North, and the hilly regions of northeast. We can thus divide the set of patterns into three families:
\begin{itemize}
\item {\bf Family 1:} patterns $1$, $2$ and $3$;
\item {\bf Family 2:} patterns $4$, $6$, $8$ and $10$;
\item {\bf Family 3:} patterns $5$, $7$ and $9$.
\end{itemize}

It is noteworthy that some locations in the northwest and southeast do not show significant activity in any of these prominent patterns. These are the regions that tend to remain dry during the monsoon. The rare rainy days in those regions are covered by the non-prominent patterns. On the other hand, the western coast receives rainfall in almost all the patterns except for $1$ and $2$. All these variations can possibly be attributed to the local orographic features.

\begin{figure}[t!]
	\centering
	\includegraphics[width=5cm,height=4cm]{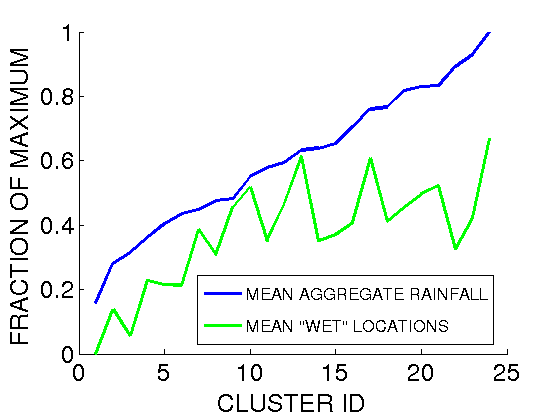}\includegraphics[width=5cm,height=4cm]{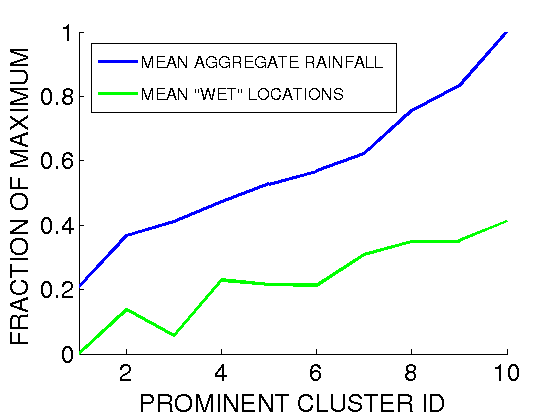}\includegraphics[width=5cm,height=4cm]{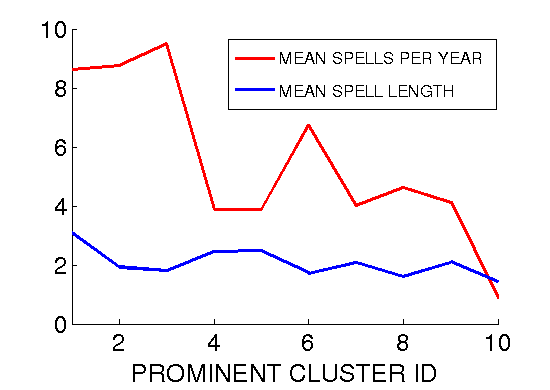}
	\caption{Properties of the clusters/patterns identified by the proposed model from the period 2000-2007. Left: mean daily aggregate rainfall for the days assigned to each cluster, and mean fraction of locations that are ``wet" on such days. Middle: same analysis for the 10 prominent patterns mentioned above. Right: Mean number of ``spells" of each pattern per year, and mean length of such spells.}
\label{fig:clusprop} \end{figure}

We now summarize few inferential observations in support of the spatial patterns obtained using our model.

\begin{figure}[t!]
\centering
\includegraphics[width=8cm,height=4cm]{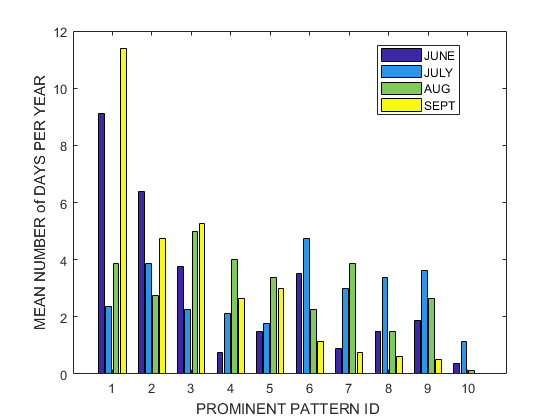}\includegraphics[width=8cm,height=4cm]{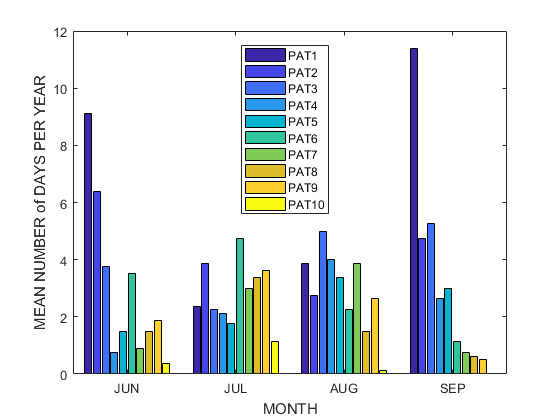}
\caption{Left: Average number of days under each prominent pattern that belong to the 4 monsoon months (based on the period 2000-2007). Right: Average number of days in each of the 4 monsoon months that were assigned to each prominent pattern (based on the period 2000-2007).}
\label{fig:clusdays} \end{figure}

\begin{itemize}
\item {\bf Aggregate rainfall and the patterns:} Observe that each spatial pattern corresponds to a distribution over daily aggregate rainfall $Y$. For any fixed $k$ belonging to the range of $U$, we consider the mean aggregate rainfall in the days associated with the corresponding cluster - \begin{equation}\label{eqn:cluster-mean}\mu_k=\text{mean}\{Y_t: U(t)=k\},\end{equation} and in Figure~\ref{fig:clusprop} we plot $\mu_k$ for each of the $24$ clusters obtained by setting $\zeta=9$ (see Table 1 of \cite{mitra2018monsoon1}), and also for the $10$ prominent clusters among them. Notably, these patterns have been sorted in ascending order of daily aggregate rainfall.
Notice that each discretized spatial pattern $\phi^d_u$ has a fraction of locations with the $Z$ values equaling the ``wet", or high rainfall state. In Figure \ref{fig:clusprop} we plot this fraction for each pattern. We argue later in Section 4 that some of these patterns are associated with ``active spells" and some with ``break spells". 
\item {\bf Patterns over the monsoon months:} Given the prominent spatial patterns, it is natural to study the distribution of these patterns across the four months (June--September) of each monsoon season. In Figure \ref{fig:clusdays}, we show the average number of days each cluster/pattern is detected in each month. Note that the first prominent pattern which has very low mean aggregate and mean number of wet locations (See Fig. \ref{fig:clusprop}) appears primarily in June and September, which can be attributed to the onset and retreat of monsoon, respectively. The patterns which are associated with high aggregate rainfall and high fraction of wet locations (see Fig. \ref{fig:clusprop}) occur mostly in July and August.
\end{itemize}

\subsection{Testing the robustness of patterns} \label{subsec:robustness}
The CDPs and CRPs shown in Figure~\ref{fig:mrf-cdp-crp} have been computed by the model for the period 2000--2007. However for the model to be robust, it should be possible to approximate $x(\cdot,t)$ and $Z(\cdot,t)$ from outside this period using the same set of patterns. In the companion paper \cite{mitra2018monsoon1} (Section 3.2), we have already established that our patterns achieve this, using the $\ell_2$ distance to compare daily rainfall data for all spatial locations to CRPs $\{\phi_u\}_{u= 1}^K$, and Hamming distance to compare $Z(\cdot,t)$ with the CDPs $\{\phi^d_u\}_{u= 1}^K$.

\begin{figure}[t!]
	\centering
	\includegraphics[width=5cm,height=4cm]{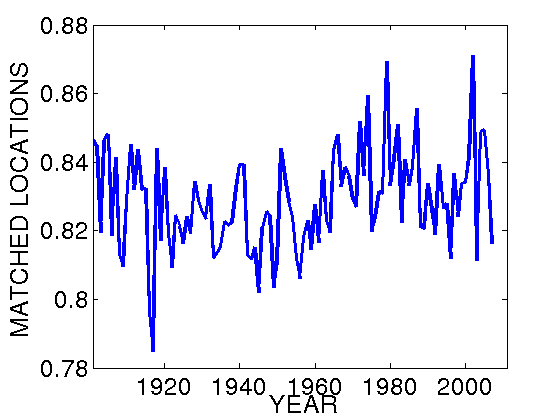}\includegraphics[width=5cm,height=4cm]{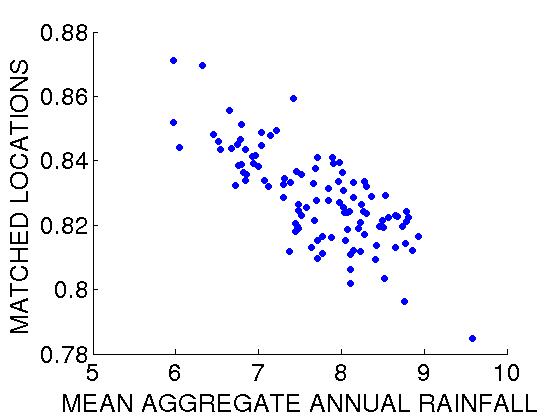}\includegraphics[width=5cm,height=4cm]{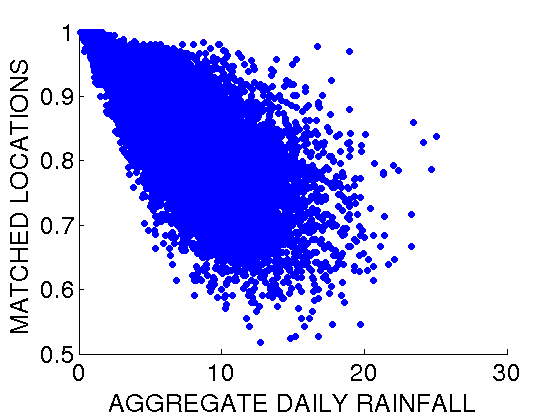}
	\caption{Hamming similarity (fraction of locations matched) between a day's DDV and the corresponding CDP varies from year to year (left panel), but the variation is not strong. But there is a clear negative correlation between annual aggregate rainfall and daily mean Hamming similarity in each year (middle panel), and this correlation holds at daily scale also (right panel).}
\label{fig:hamm-sim} \end{figure}

In this part, we examine how such fitting of patterns varies from year to year. We compute the mean Hamming similarity\footnote{Hamming similarity between two binary sequences $a_1$ and $a_2$ of length $n$, can be defined as $\frac1n\times (n-\text{Hamm}(a_1,a_2))$, where $\text{Hamm}(a_1,a_2)$ is the usual Hamming distance between the two sequences $a_1$ and $a_2$, i.e. number of elements where they are different.} between each day's discrete state vector $Z(\cdot,t)$ and the corresponding CDP for each year in the period $1901-2007$, and the results are displayed in the left part of Figure \ref{fig:hamm-sim}. This figure plots the mean Hamming similarity index between $Z(\cdot,t)$ and the corresponding CDP, for each of the $107$ years, from $1901$ to $2007$. We find that this quantity does not change very significantly across the years, and it is not strikingly different in years inside and outside the model training period of $2000-07$. During the period $2000-07$ the mean daily Hamming similarity index is 0.84 (the CDP and $Z(\cdot,t)$ are same in 299 of the 357 locations on average) while in the period 1901-1999 the same is 0.83 (296 of the locations).

When comparing the mean Hamming similarity index to mean aggregate rainfall  of various years, it is noticeable in the scatter plot of Figure \ref{fig:hamm-sim} that the two variables exhibit a negative correlation. The patterns seem to be fitting relatively poorly to the binary fields $Z(\cdot,t)$ in those years that the annual aggregate rainfall is high. This can possibly be explained by the observation that such years are likely to have a larger number of days with excessive rainfall, and our patterns do not fit well on such days, which is seen in the right part of Figure \ref{fig:hamm-sim}, where we see a clear negative correlation between the spatial aggregate rainfall on a day and the Hamming similarity between that day's assignment of the $Z$ variables and the corresponding CDP.

The above observation leads us to further investigate the relationship between distribution of patterns across days and the aggregate rainfall of each year, prompting us to identify the patterns which are more prominently seen in drought, and excess years\footnote{We identify the mean $(\mu)$ and standard deviation $(\sigma)$ of annual aggregate rainfall, and follow the Indian Meteorological Department's definitions~\cite{rajeevanspells} of excess and deficient rainfall years as those getting more rainfall than $\mu+\sigma$ and less rainfall than $\mu-\sigma$ respectively}. 

\begin{figure}[t!]
	\centering
	\includegraphics[width=6cm,height=4cm]{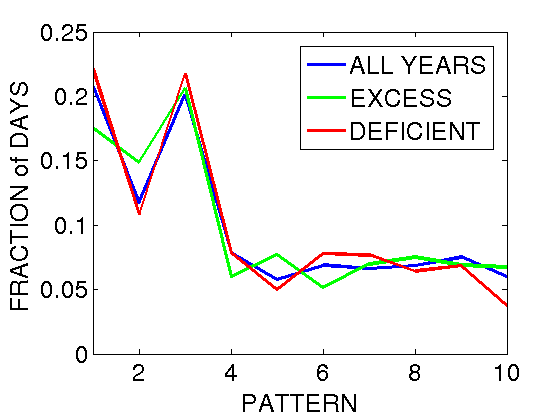}\includegraphics[width=6cm,height=4cm]{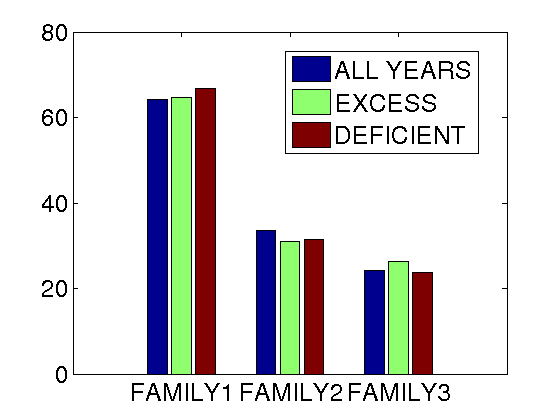}
	\caption{Fraction of days assigned to different patterns (left) and different families (right); also computed separately for years of excess and deficient all-India aggregate rainfall}
\label{fig:frac-days} \end{figure}

In the left panel of Figure \ref{fig:frac-days}, we plot the mean number of days assigned to each of the patterns in deficient, excess and all years (in the period 1901-2007) separately, whereas on the right panel, we show the fraction of days assigned to each of the three families 1, 2 and 3 in deficient and excess rainfall years separately. We observe that the fraction of days assigned to the family 1, which corresponds to dry patterns, is highest in the deficient-rainfall years as expected. In contrast, in the excess-rainfall years, we find the fraction of days assigned to family 3 (Central India and Western coast) is highest. In such years, the fraction of days assigned to Family 2 (rainfall over Gangetic plain) is less compared to normal.

\subsection{Seasonal evolution of spatial patterns}\label{subsec:time-evolve}
A key aspect of Indian monsoon is its intra-seasonal oscillation. The rains do not remain confined to one region, but propagate across the landmass. The area covered by rainfall expand and contract and shift as the season proceeds. One key motivation of creating the discretized view in this work is to observe these propagation patterns. 

%\begin{figure}[t!]
%\centering
%\includegraphics[width=6cm,height=6cm]{transmat_full1.png}\includegraphics[width=6cm,height=6cm]{transmat_107yrs.png}
%\caption{Transition matrix of the patterns between consecutive days. In the left matrix the diagonal elements dominate, indicating strong tendencies of self-transition in each state. In the right matrix, the diagonal elements have been set to 0 to highlight the transitions other than same-state transitions.}
%\label{fig:transmats}
%\end{figure}

\begin{figure}[t!]
\centering
\includegraphics[width=0.32\columnwidth]{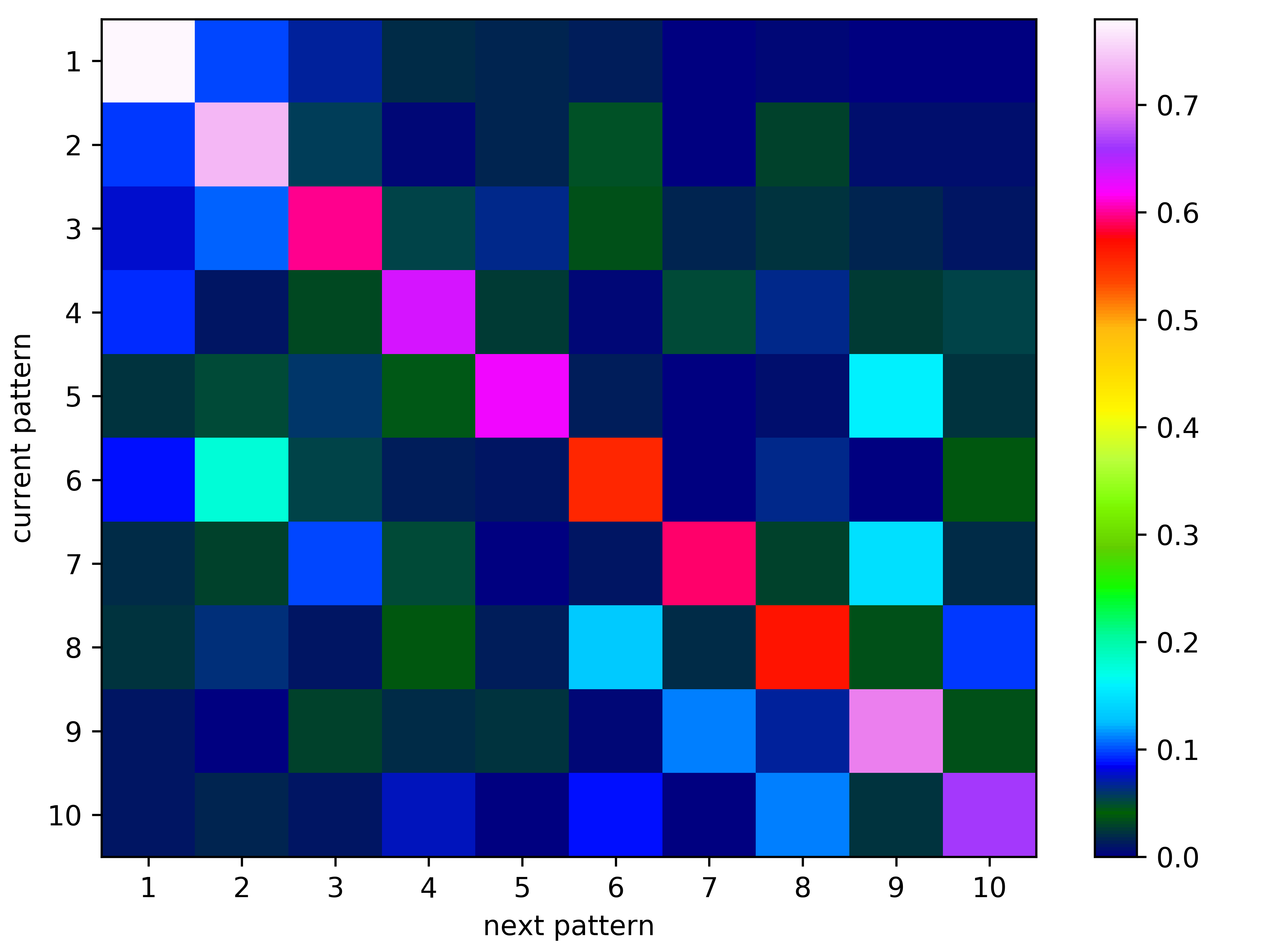}
\includegraphics[width=0.32\columnwidth]{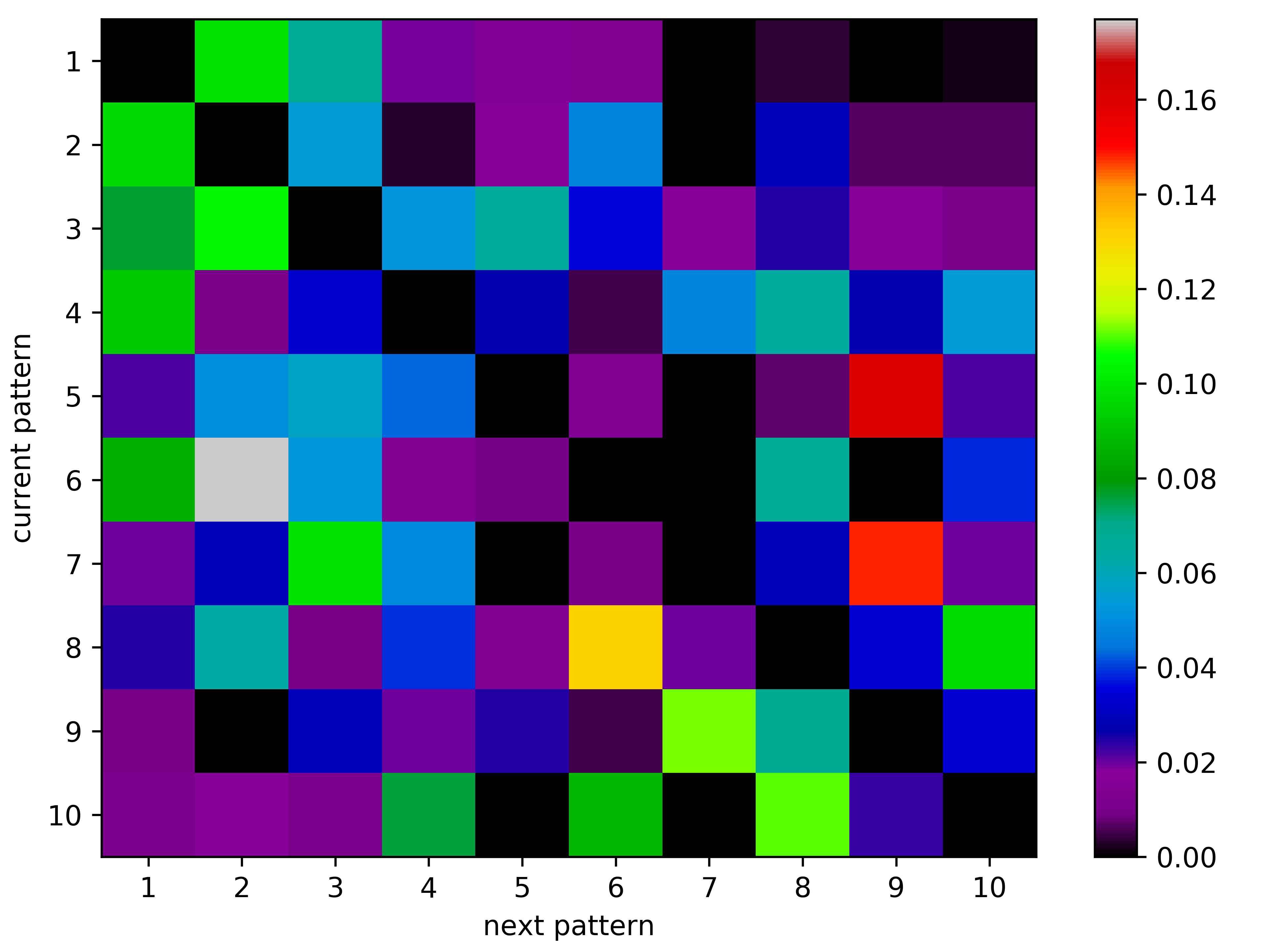}
\includegraphics[width=0.32\columnwidth]{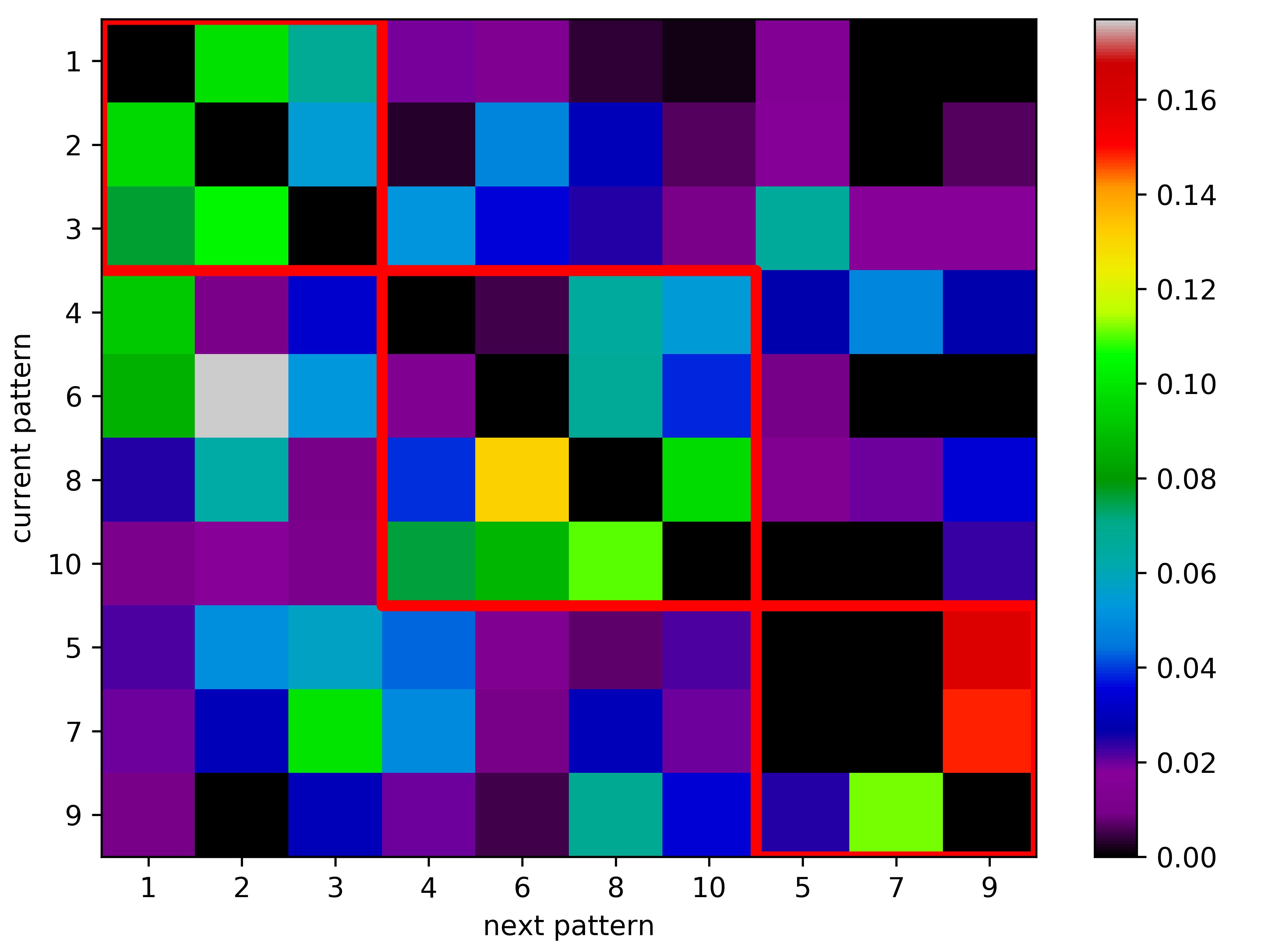}
\caption{Transition matrix of the patterns between consecutive days. In the left matrix the diagonal elements dominate, indicating strong tendencies of self-transition in each state. In the middle matrix, the diagonal elements have been set to 0 to highlight the transitions other than same-state transitions. In the right matrix, the states have been rearranged according to Families 1,2,3 to highlight the block-diagonal nature of the matrix. Each block along the diagonal represents a family, and we see that intra-family transitions are more frequent than inter-family transitions.}
\label{fig:transmats} 
\end{figure}

\subsection{Probability distribution of prominent spatial transitions} 
We first study how the spatial distribution of rainfall changes between consecutive days. The simplest way to study this is the \emph{state transition matrix} which encodes the distribution $\mathbb{P}(U(t+1)=k|U(t)=l)$ where $k,l\in\{1,10\}$, indices of the prominent patterns. We make a maximum-likelihood estimate of this quantity using the $U$-variable for each day inferred by our model, and the matrix is plotted in the left panel of Figure \ref{fig:transmats}. Clearly the diagonal terms dominate, which means that once a pattern appears, it is likely to persist for a few days. This is in agreement with the right panel of Figure~\ref{fig:clusprop}, which plots the mean spell length of each pattern. In order to  highlight transitions to other states we have set the probability of self transitions (diagonal elements of the transition matrix) to zero in the right panel of Figure~\ref{fig:transmats}. The reader may observe that with regards to the families of patterns described earlier, the intra-family transitions appear more likely than the inter-family transitions.

\begin{figure}[t!]
\centering
\includegraphics[width=\columnwidth]{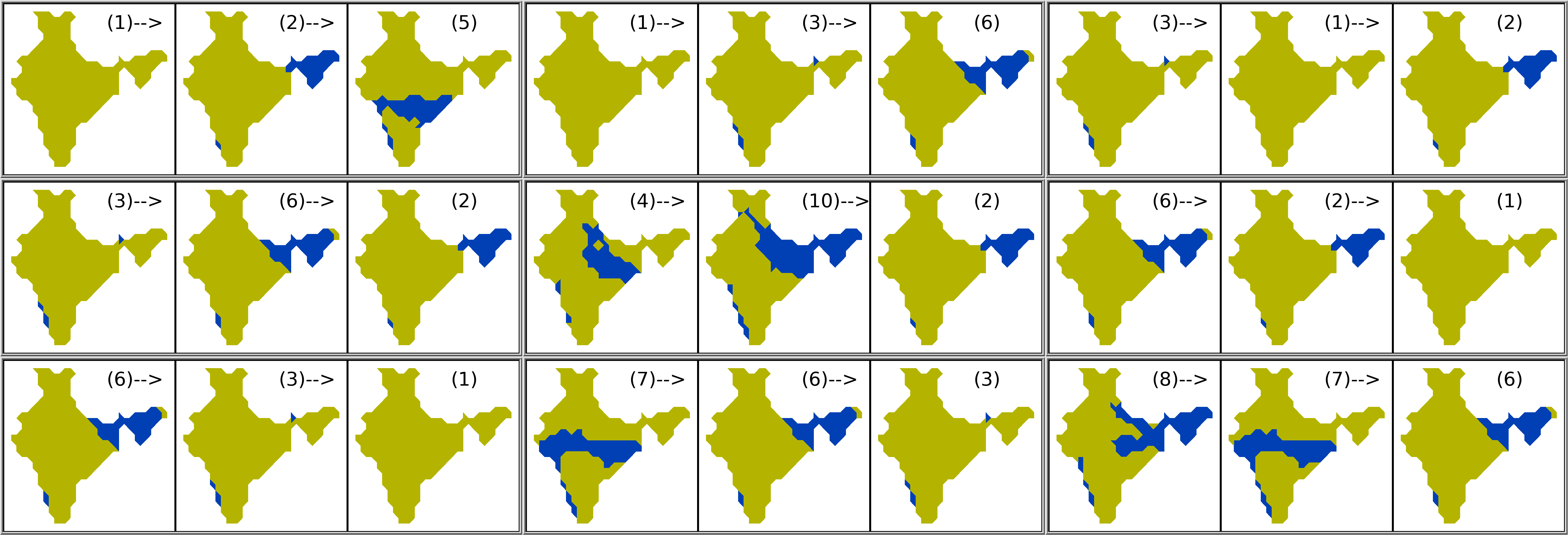}
\caption{Frequent 3-step transition patterns among the prominent Canonical Discrete Patterns identified by the proposed model and displayed in Figure~\ref{fig:mrf-cdp-crp}} \label{fig:transpatt} \end{figure}

\noindent{\bf Spatial transition patterns:} From the sequence of monsoon days in the period 2000-20007 that we used to find the canonical patterns by the proposed model, we attempted to find frequent subsequences of the canonical discrete patterns. For the sake of simplicity, we studied subsequences of length $3$. However, since self-transitions are the most frequent we collapse such transitions while identifying the $3$ subsequences. For instance, a sequence of $\{1 3 3 4 4 4 5\}$ was identified as $\{1 3 4 5\}$. Then, the frequency of such transitions is computed, and the most frequent $3$-subsequences are are shown in Figure \ref{fig:transpatt}. It is not difficult to observe that most of the subsequences shown indicate advance or retreat of the rainy areas, i.e. transitions into/out of active and break spells. The spatial continuity of the transitions is quite noteworthy. 

\subsubsection{Self transitions: length of spells}\label{subsubsec:self-transitions} The self-transitions define ``spells" where the same pattern repeats for a few days in succession. Such self-transitions are studied separately in Figure~\ref{fig:clusprop}, where we plot the mean spell length for each prominent pattern, and also the mean number of spells that occur per season. It may be noted that the driest pattern $1$ has the longest mean spell length and also high number of spells per season, while wettest pattern $8$ has fewest mean number of spells per season.

The self-transitions are closely related to the property of \emph{temporal coherence}, which has already been discussed. This property states that the spatial pattern of rainfall on any day is likely to be similar to the previous day's, and hence successive days are likely to be associated with the same pattern. This property is not enforced by any of the clustering approaches considered. Our model enforces temporal coherences at local scales through $Z$-variables which indirectly affects the $U$ variables. It turns out that the proposed model shows self-transition of $U$-variable on $518$ of the $975$ occasions in the considered period. In contrast, the numbers of self-transitions are only $381$, $271$ and $294$ in case of K-means, Spect1 and Spect2 respectively. This shows that the proposed model based on Markov random field can preserve the temporal coherence property of the system.

%%%%%%%%%%%%%%%%%%%%%%%%%%%%%%%%%%%%%%%%%%%%%%%%%%%%%
%%%%%%%%%%%%%%%%%%%%%%%%%%%%%%%%%%%%%%%%%%%%%%%%%%%%%

\section{Results: identifying the active and break spells} \label{sec:actbrk}
In this section, we study various characteristics of the active and break spells at local and all-India scales using the allocations of $(Z,U,V)$ variables to various temporal and spatial states. 

\subsection{All-India Active-Break spells} \label{ssec:actbrk}
As already stated, a very important feature of Indian monsoon is the existence of ``active spells" and ``break spells". Such a spell of days may be characterized by certain rainfall patterns, along with high or low aggregate rainfall. According to most conventions, such as~\cite{rajeevanspells}, these days are identified, for the whole Indian landmass, by comparing the daily aggregate rainfall against some threshold. Unlike in~\cite{rajeevanspells} where the authors considered aggregate rainfall over the monsoon zone during July and August, we consider aggregate rainfall over all of India during June-September, hence including the onset period within our ambit of study. Following the convention of~\cite{rajeevanspells} the aggregate daily thresholds for active and break days are given by $(\mu_Y+\sigma_Y)$ and $(\mu_Y-\sigma_Y)$, respectively, where $\mu_Y$, $\sigma_Y$ are the mean and standard deviations of $Y$, computed over the entire period of $976$ days. An ``active day" is a day when the daily aggregate rainfall $Y(t)$ satisfies $Y(t)\geq\mu_Y+\sigma_Y$, and an ``active spell" is then defined by a run of $3$ or more active days.  A ``break day" is similarly defined as a day of aggregate rainfall below $(\mu_Y-\sigma_Y)$, and ``break spell" by a run of $3$, or more break days.  We denote the set of active and break days identified this way as $ACT_0$ and $BRK_0$, respectively.

From the temporal clusters of spatial patterns obtained using our proposed method, we have already seen in Figure~\ref{fig:clusprop} that different prominent patterns are associated with different rainfall patterns, and thus each cluster has different values of mean daily aggregate rainfall, $\mu_k$ (defined in \eqref{eqn:cluster-mean}). We observe that among the 10 prominent patterns (see Table 3 of the companion paper~\cite{mitra2018monsoon1}), $4$ prominent clusters, satisfy $\mu_k\geq\mu_Y+\sigma_Y$, and $1$ prominent cluster satisfies $\mu_k\leq\mu_Y-\sigma_Y$, where $\mu_k$ is the mean aggregate rainfall for the $k$-th cluster (see \eqref{eqn:cluster-mean}). We can call these clusters as ``active cluster" and ``break cluster", respectively. Since each day is assigned to one single cluster through the variable $U$, we thus have an alternative method of identifying active and break spells. A day belonging to an active cluster may be considered an active day, and a day belonging to a break cluster may be called a break day. This is realistic due to the observation in the companion paper~\cite{mitra2018monsoon1} (Table 3) that these clusters exhibit homogeneity with respect to daily aggregate rainfall, as they have low intra-cluster variation. Let us denote this new set of active/break days as $ACT_1$ and $BRK_1$, respectively, and the corresponding runs of $3$ or more such days as the active and break spells, respectively.

%{\color{red}We now investigate if the information drawn from our proposed model can be used to theorize other ways to characterize such spells. The basic premise underlying the following method is the uniformity within a given temporal cluster formed by the temporal cluster identifier $U$. For any fixed $k$ belonging to the range of $U$, we consider the mean aggregate rainfall in the days associated with the corresponding cluster - $\mu_k=\text{mean}\{Y_t: U(t)=k\}$, and the corresponding standard deviation - $\sigma_k = \text{s.d.}\{Y_t: U(t)=k\}$. Then, for any given day $t$, let $k'$ be its temporal cluster assignment. We test if $Y_t$ for the given day $t$ is more than one $(\mu_{k'}+\sigma_{k'})$. If the above procedure results in affirmative for three consecutive days then we call it an active spell according to our procedure. We shall identify the set of all such active spells by $ACT_1$. We shall similarly define the break spells $BRK_1$ by testing if $Y_t$ for a given day $t$, is smaller than $(\mu_{k'} - \sigma_{k'})$, where $k'$ is the temporal cluster assignment given to day $t$.} 

\begin{figure}
\centering	%\includegraphics[width=15cm,height=6cm]{ACTBRK.png}
\includegraphics[width=\textwidth]{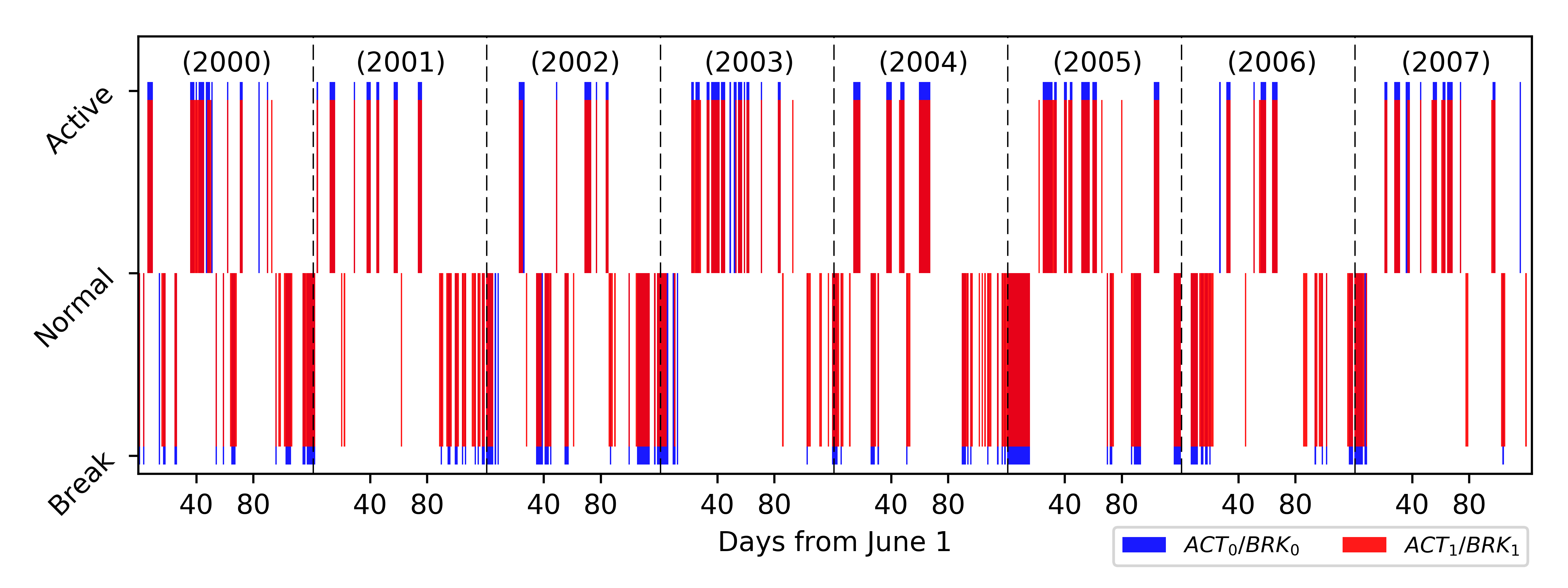}
\caption{The bottom panel shows active and break days according to the method proposed in this paper, i.e. $ACT_1$ and $BRK_1$, and the top panel shows those according to the method of~\cite{rajeevanspells}, i.e., $ACT_0$ and $BRK_0$}
\label{fig:ACTBRK} 
\end{figure}

A comparative chart is shown in Figure~\ref{fig:ACTBRK}, where it is evident that most of the active and break spells appear common to both the criteria. In fact, our analysis shows that there are $152$ days belonging to both $ACT_0$ and $ACT_1$, whereas individually, the number of days belonging to $ACT_0$ and $ACT_1$ are $161$ and $169$, respectively. Similarly, there are $145$ days common to $BRK_0$ and $BRK_1$, whereas there are $153$ days in $BRK_0$, and $214$ days in $BRK_1$. On closer inspection, we observe that the mean aggregate rainfall associated with $ACT_0$ is $12.8$ mm/day/grid, and is $12.55$ mm/day/grid for $ACT_1$. On the other hand, the mean number of locations receiving above-mean rainfall in the days of $ACT_1$ is higher than that for $ACT_0$. Therefore, $ACT_1$ identifies days where not only is the aggregate rainfall high, but in addition the rain is well-distributed also. In case of break spells, the days in $BRK_0$ have less rainfall than those of $BRK_1$ and fewer locations have above-mean rainfall. This is because the lone cluster associated with $BRK_1$ has a very high number of days associated with it when compared with those obtained by way of  thresholding. The clusters of $U$ in case of the model were formed based not only on $Y$ but also on the spatial patterns. 
% It turns out that ``active days" have multiple spatial patterns associated with them, but ``break days" do not. {\color{red}{I didn't fully understand this statement. Can this be made clearer?}}

We also report some of the statistics of the new set of active and break spells, as identified by $ACT_1$ and $BRK_1$ for the period $2000-07$. In this convention, the mean length of active spells is $4.37$, compared to $4.0$ in case of $ACT_0$. Also, in case of $ACT_1$ there are $32$ active spells in the period as compared to $29$ for $ACT_0$. There are also $33$ break spells in the given period with mean length $5.12$ in case of $BRK_1$, compared to just 18 spells with mean length of $5.33$ for $BRK_0$.

\subsection{Local Wet-Dry Spells} \label{ssec:locwetdry}
As against defining spells at all-India scale, it is pertinent to define such rainfall patterns over smaller scales. In ~\cite{singhspells}, the authors attempt to define and study such spells at scales smaller than all-India, but their study is limited to certain zones defined by the Indian Meteorological Department. In this section, we shall consider wet and dry spells at grid-scale, based on the discrete states assigned by $Z$. 

We argue that the proposed model is better at allocating the discrete states to rainfall measurement stations than methods based on deterministic fixed thresholds, at the local or global level, as the proposed model maintains spatio-temporal coherence of the discrete states unlike methods based on hard thresholds.
%\sout{We argue that deciding the discrete states based on the proposed model is better than doing so based on local or global thresholds, because the proposed model is able to maintain spatio-temporal coherence of the latent states, unlike the latter approach. }
In case of the proposed model, on average $88\%$ of neighboring grid-pairs have the same discrete state on any day, as opposed to $83\%$ when using local daily means as the threshold. Also, in case of the proposed model any location remains in same state as the previous day on $92\%$ of days, while in case of the location specific threshold approach this number is only $77\%$. Using a location independent threshold like $5$ mm results again in low temporal coherence. Clearly the proposed approach can help us identify more coherent local conditions, which is important because rainfall is a disorderly outcome of coherent and large-scale climatic systems.

Using the discrete states {assigned by the model ($Z$), we identify wet and dry spells at each of the $357$ grid-locations. At each location, $Z$ takes two values: wet and dry. A continuous sequence of days assigned to the wet state forms a wet spell, while such a continuous sequence assigned to the dry state forms a dry spell.  On average, a wet spell lasts $5.6$ days, while a dry spell lasts $23.2$ days. However, the mean lengths of wet and dry spells vary across locations, and the grid-wise mean lengths of wet and dry spells are shown in Figure \ref{fig:grid-spell-len}.

\begin{figure}
	\centering
    \includegraphics[width=\columnwidth]{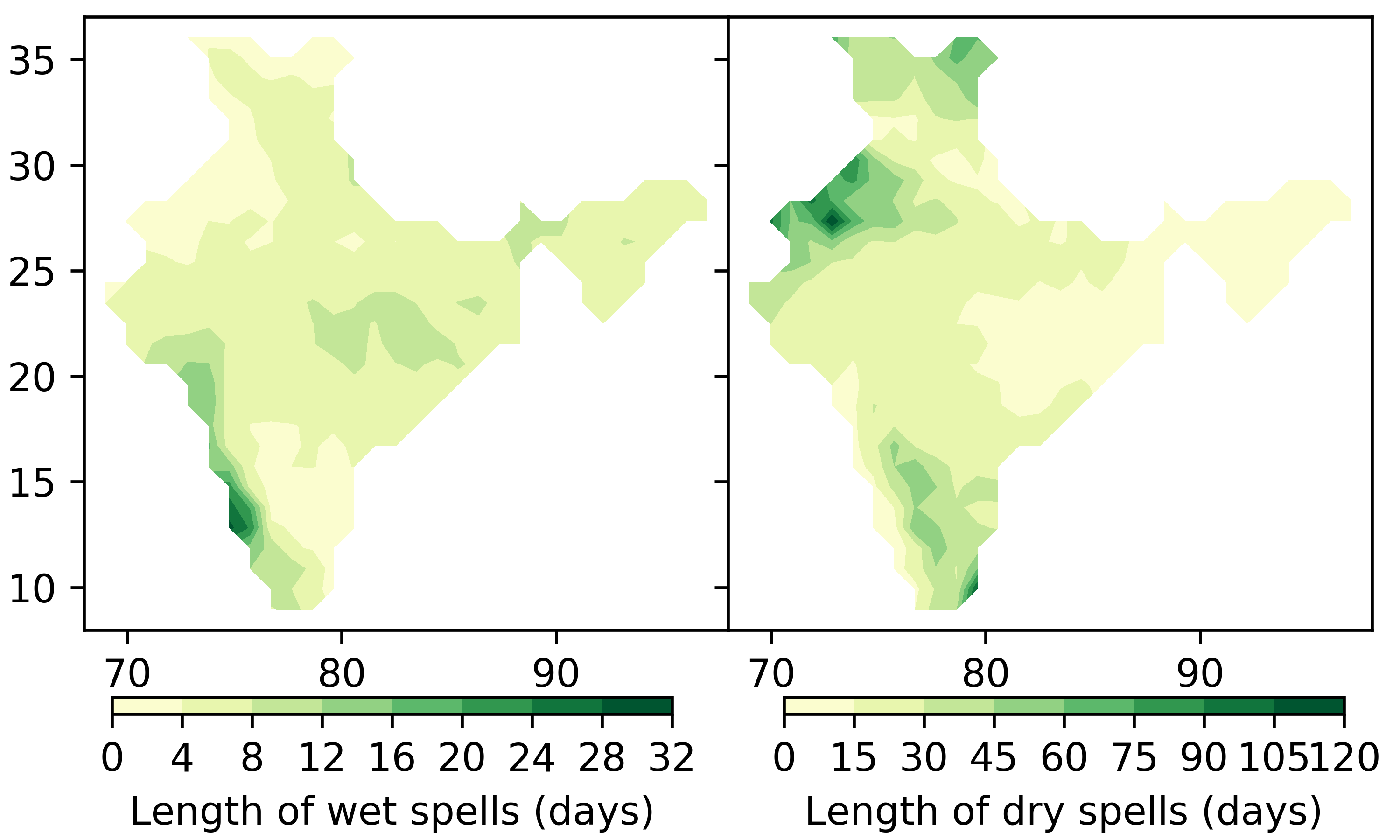}
	\caption{Mean lengths of wet spells (left) and dry spells (right) at different grids. The spells are identified according to the discrete state assignments by the proposed model, as discussed in section~\ref{ssec:locwetdry}}
\label{fig:grid-spell-len} \end{figure}

\subsection{Regional Wet-Dry Spells} \label{ssec:regwetdry}
We next turn to defining wet-dry spells at an intermediate scale. This is motivated by the observation that several grid-points which are typically adjacent to each other tend to have their local wet and dry spells simultaneously. So it may be possible to partition the landmass into regions consisting of a few grid-points (probably geographically adjacent) each of which act as ``coherent units". The Indian Meteorological Department (IMD) already has divided the landmass into $19$ ``homogeneous regions", while other works such as~\cite{eofcluster,annualcluster} identify clusters of grids based on time-series of daily or annual rainfall, and both use Empirical Orthogonal Functions for this purpose. However clustering algorithms like K-means and Spectral Clustering are better placed to solve this problem. Such clustering algorithms usually suffer the drawback of having to specify the number of clusters \emph{a priori}.

The model proposed in this paper can identify clusters of locations through the variable $V$ based on the local discrete states $Z$, which encode information about local wet- and dry- spells. The grid-points in the same temporal clusters are likely to have the same discrete states on most days and hence have their dry and wet spells simultaneously. Spatial Coherence of clusters is not enforced directly on the $V$-variables in the model, but indirectly through the $Z$-variables which are spatially coherent due to the MRF potential functions. Indeed, Figure~\ref{fig:regional-spell-length} shows strong spatial coherence in the mean lengths of grid-specific wet and dry spells. The number of temporal clusters does not have to be specified by the user, but the parameter $\zeta$ provides an indirect handle, like the $\eta$-parameter for the spatial clusters encoded by $U$. Smaller values of $\zeta$ implies the formation of more clusters, and this is shown in Table 1 along with measures of accuracy.

As in the case of daily clusters (Section \ref{sec:prominent-sp-patterns}), we evaluate the temporal clustering using intra-cluster similarity. We use the same baselines as those used to evaluate the clusters in the companion paper~\cite{mitra2018monsoon1}: K-Means~\cite{kmeans} where the real-valued rainfall time-series $x(s,\cdot)$ are clustered; Spectral Clustering~\cite{spect} based on Euclidean distance between the rainfall time-series (Spect1); and Spectral Clustering based on Hamming Distance (Spect2) between discretized time-series $Z(s,\cdot)$. In case of Spect2, this discretization is done by comparing local daily rainfall $X(s,t)$ with local daily mean $\mu_s$. For each cluster we identify the canonical time-series (CTS) $\theta$ and the canonical discrete series (CDS) $\theta_d$ analogous to the canonical rainfall patterns (CRP) and canonical discrete patterns (CDP) in case of daily clusters. The evaluation criteria are also the same as in case of daily clusters in the companion paper~\cite{mitra2018monsoon1}(Table 4) - intra-cluster standard deviation of total rainfall at each location (across the entire period of $976$ days), $\ell_2(\theta)$: the mean $\ell_2$ (Euclidean) distance between the time-series at each grid (RTS) and the corresponding cluster's canonical time-series (CTS), and $\text{Hamm}(\theta_d)$: the mean Hamming distance between the discrete time-series at each grid (DTS) and the corresponding cluster's canonical time-series (CDS). The results are shown in Table 1.

\begin{scriptsize}
	\begin{table}
		\begin{tabular}{| c | c | c | c | c | c | c | c | c | c |}
			\hline
			$\zeta$ & \multicolumn{3}{| c |}{std(YY)} & \multicolumn{3}{| c |}{$\ell_2(\theta)$} & \multicolumn{3}{| c |}{Hamm($\theta_d$)}\\
			\hline
			(\#clusters) & MRF & KMeans & Spect1 & MRF & KMeans & Spect1 & MRF & KMeans & Spect2\\
			\hline
			10 (60)  & 1.23 & \textbf{0.98} & 1.3 & 279 & \textbf{268} & 290 & \textbf{73} & 151 & 114\\
			12 (51)	 & 1.31 & \textbf{1.18} & 1.39& 291 & \textbf{277} & 305 & \textbf{81} & 161 & 121\\
			15 (40)  & 1.29 & \textbf{1.27} & 1.55& 311 & \textbf{298} & 324 & \textbf{95} & 174 & 131\\
			20 (29)  & 1.81 & 1.64 & \textbf{1.57}& 345 & \textbf{323} & 345 & \textbf{108}& 191 & 135\\
			25 (24)  & \textbf{1.67} & 1.90 & 1.84& 367 & \textbf{347} & 362 & \textbf{124}& 203 & 144\\
			\hline 
		\end{tabular}	
		\caption{Comparison of daily cluster properties, by varying the number of clusters through $\zeta$ parameter of the proposed model. $std(YY)$ is the average intra-cluster standard deviation of aggregate rainfall at each location $YY$, $\ell_2(\theta)$ is the mean $\ell_2$-distance of RTSs to CTS $\theta$ of corresponding cluster and $Hamm(\theta_d)$ is the mean Hamming distance of DTSs to CDS $(\theta_d)$ of corresponding cluster.}
\label{tab:comp-clus} \end{table}
\end{scriptsize}

These results show that the proposed model gives the best clustering results with respect to the discrete representation, i.e. the Hamming distance of each discretized time-series to the corresponding cluster's canonical discrete series. Hence this approach is the best to identify regions that have simultaneous wet and dry spells. With respect to the other two measures regarding the real-valued time-series the proposed method lags somewhat behind K-means, which is expected since the model does not consider the real-valued RTS to assign the clusters $V$ unlike K-means and Spect1. Even then, the proposed model often outperforms Spect1 and comes close behind K-means.

\begin{figure}
	\centering
	\includegraphics[width=0.45\columnwidth]{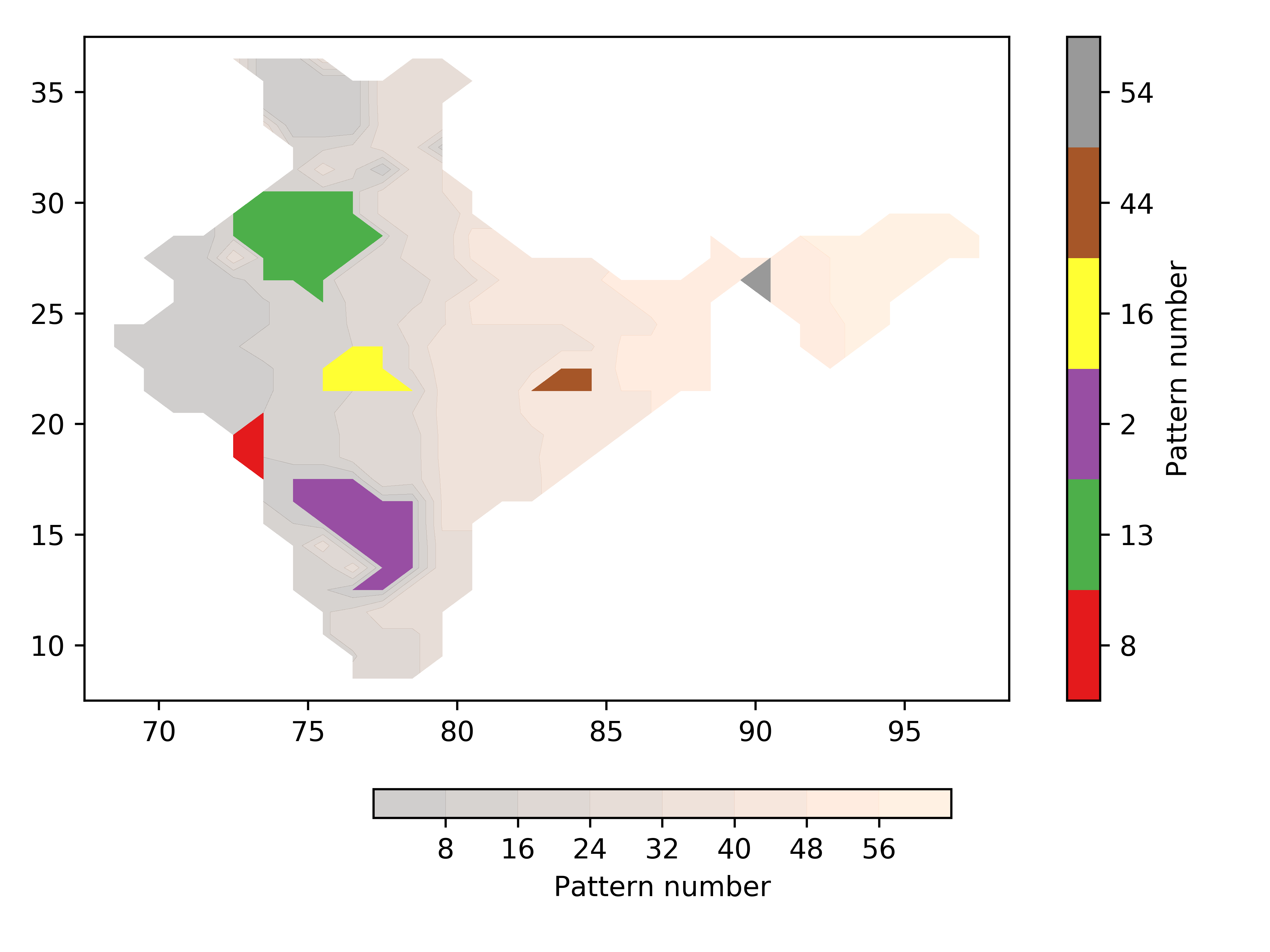}
    \includegraphics[width=0.45\columnwidth]{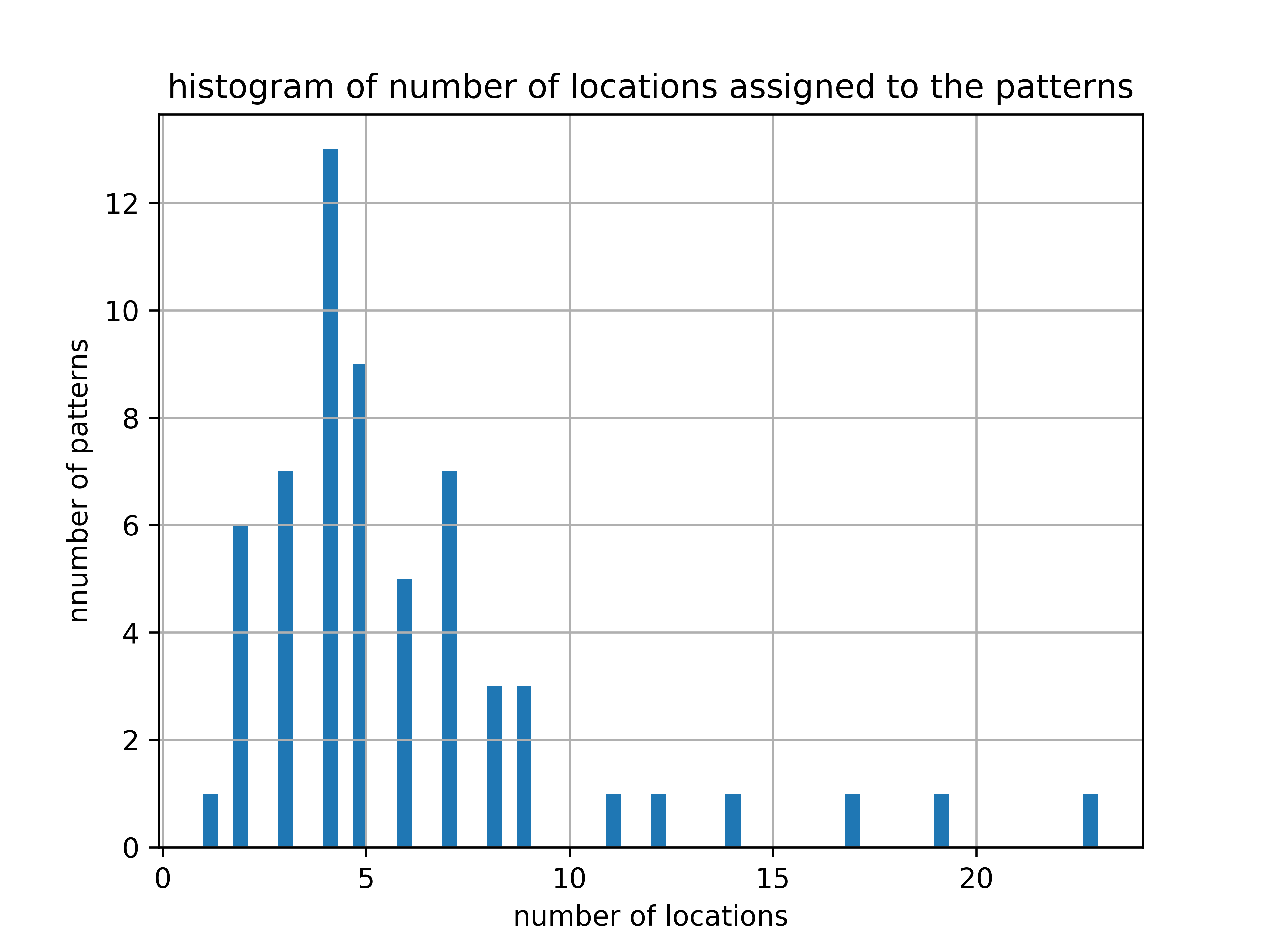}
    \includegraphics[width=\columnwidth]{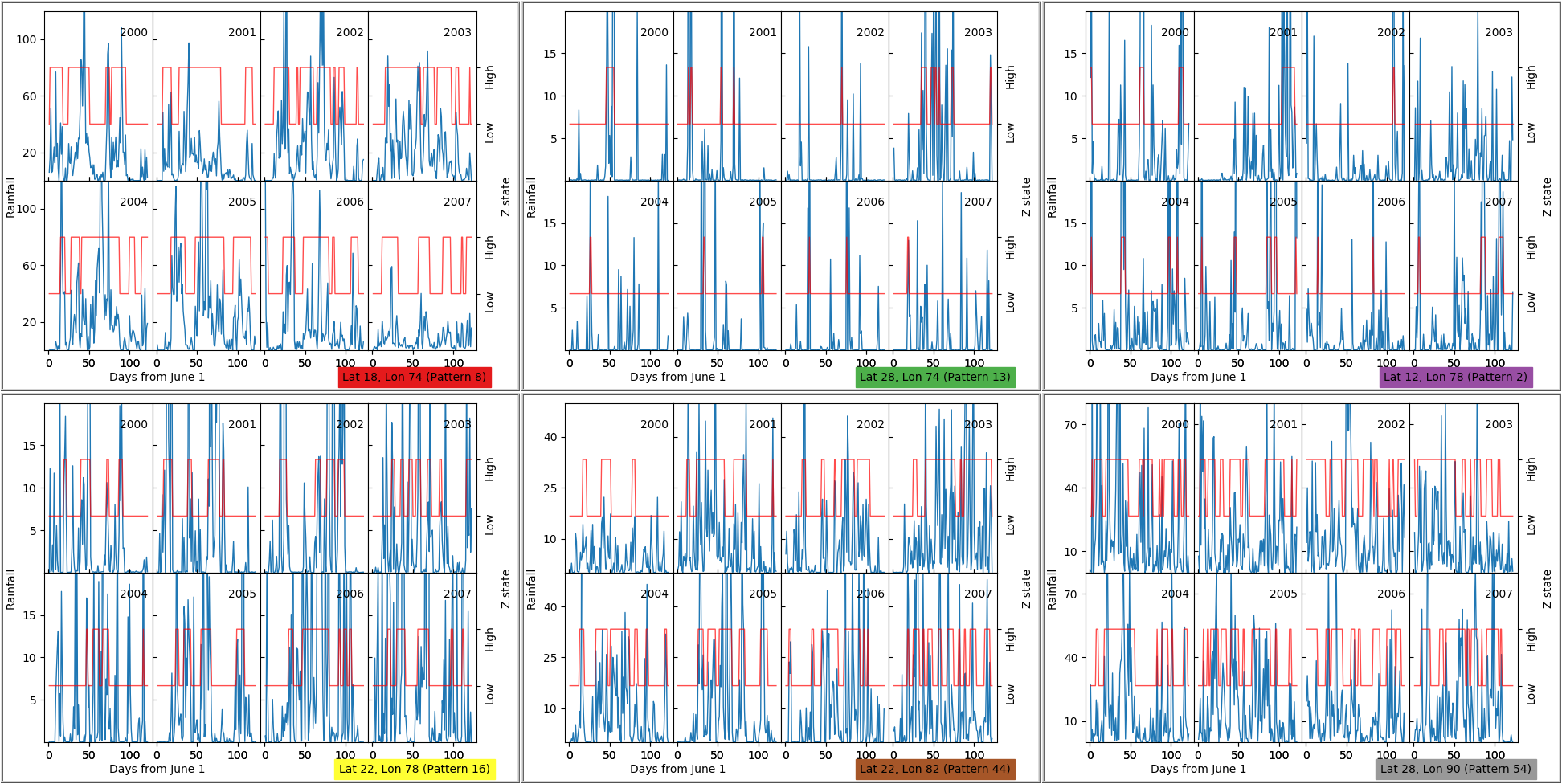}
    \caption{The 60 different spatial clusters corresponding to the 60 temporal patterns are shown in the background shading of top left panel which also highlights six illustrative clusters. The time series for the discrete variable $Z$ and for the rainfall $X$ at one of the locations in each of the six cluster are shown in the bottom panel. The histogram of the number of locations (out of a total of 357) allocated to each of the 60 clusters is shown in the top right panel.}
\label{fig:patt-loc} \end{figure}

% \begin{figure}
% 	\centering
%     \includegraphics[width=\columnwidth]{rain-Z-series-loc-montage-fr10-theta-dpi100.png}
%     \caption{same as previous fig but showing the pattern corresponding to the cluster to which the location belongs. WHICH one should we keep? {\color{red}{The temporal patterns for a location and the rainfall at that location are not quite ``matching'' very well...}}}
% \label{fig:patt-loc-alt} \end{figure}

In Figure~\ref{fig:patt-loc}, we show the spatial clusters/regions formed by setting $\zeta=10$. As shown in Table 1, this setting produces $60$ regions which are spatially coherent, i.e., each of the region is contiguous. The number of locations assigned to each of these $60$ spatial clusters varies between $1$ and $23$ with a large number of clusters ($48$) containing less than 8 locations while a small number of them ($12$) contain 8 or more positions. But only 147 locations out of a total of 357 locations are covered by these ``large'' clusters while the remaining 210 locations belong to the ``small'' clusters. Thus, in contrast to the temporal clustering where a small ($10$) number of patterns are able to cover a very large fraction of days enabling us to identify ``prominent patterns'' as explained in the companion paper~\cite{mitra2018monsoon1}, we have not identified prominent spatial clusters. The histogram on the top right panel of Figure~\ref{fig:patt-loc} shows the distribution of number of locations assigned to these patterns. The bottom panel of Figure~\ref{fig:patt-loc} also shows the CDS for some of the spatial clusters and the RTS for some locations within these clusters. These 6 selected regions have fairly diverse rainfall characteristics, as can be seen from their time-series in the figure. Clockwise from topl-left, these regions are: the rainy Western coast near Mumbai (region 8), the dry desert regions of Rajasthan (region 13), the moderately dry regions of Karnataka that lie in the rain shadow of Western Ghats range (region 2), the highly wet hilly regions of Meghalaya (region 54), the moderately wet region of Jharkhand which receives rain-clouds from the eastern coasts (region 44), and Southern parts of Madhya Pradesh which lies in the core ``monsoon zone'' and receives reasonably good rainfall(region 16). As explained in section~\ref{ssec:regwetdry}, we also consider the wet and dry spells at the scale of these regions. The mean length of wet spells is $7$ days and that of dry spells is $23$ days. However there is a lot of variation across the landmass, as illustrated on the maps in Figure~\ref{fig:regional-spell-length}.

\begin{figure}
	\centering
    \includegraphics[width=\columnwidth]{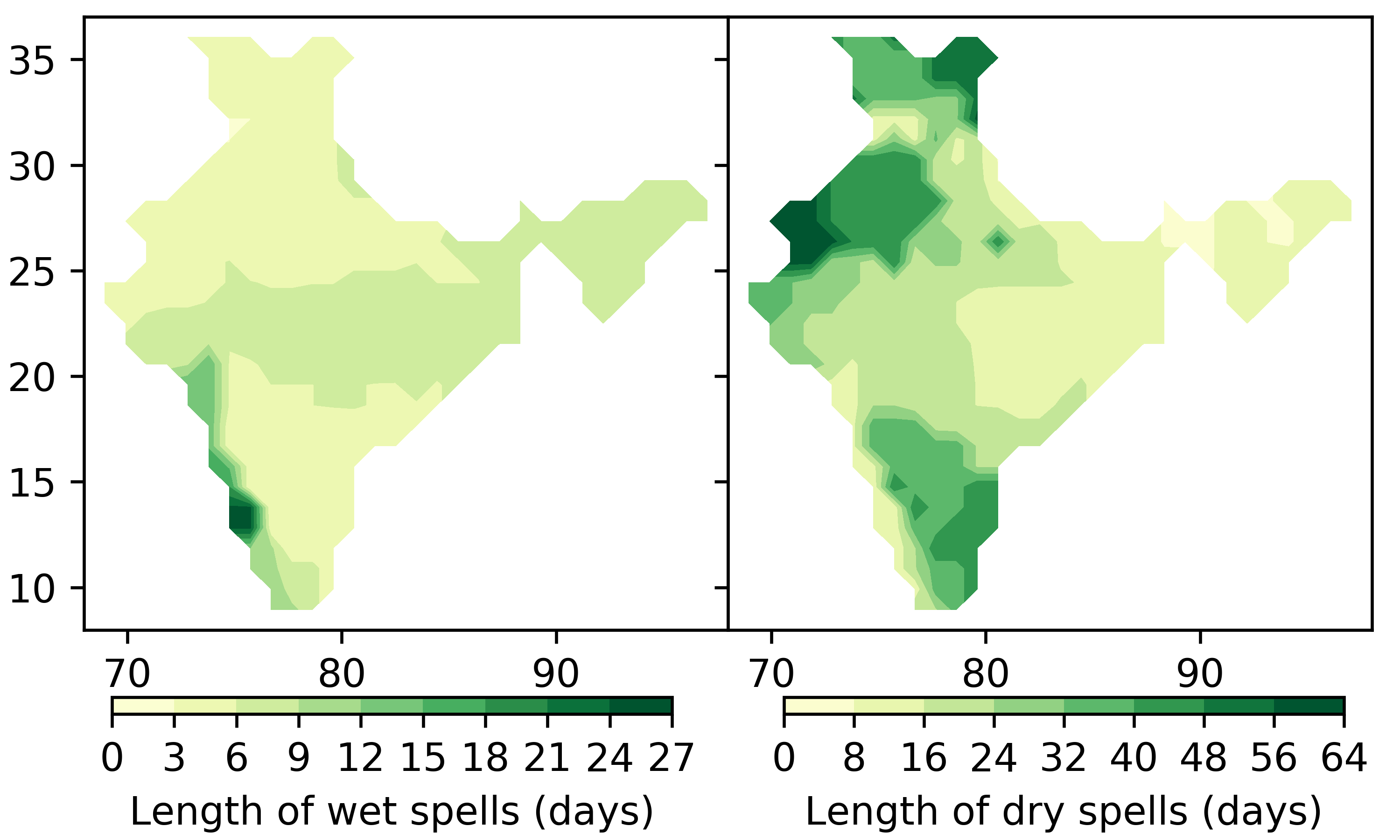}
	\caption{Mean lengths of wet spells (left) and dry spells (right) at regional scale, based on the 60 spatial clusters, as discussed in section~\ref{ssec:regwetdry}.}
\label{fig:regional-spell-length} \end{figure}

%%%%%%%%%%%%%%%%%%%%%%%%%%%%%%%%%%%%%%%%%%%%%%%%%%%%%
%%%%%%%%%%%%%%%%%%%%%%%%%%%%%%%%%%%%%%%%%%%%%%%%%%%%%
\section{Discussion and future directions}
\label{sec:discuss}
In our companion paper, we introduced a mathematical model to analyze rainfall data during the Indian monsoon season provided by Rajeevan et al. \cite{rajeevanspells}. The upshot was an allocation of discrete labels Z, representing ``wet'' or ``dry'' states, to each spatial and temporal location. In addition, we identified a number of canonical patterns of such states across all spatial locations. Each canonical discrete spatial pattern is representative of a specific rainfall pattern over the Indian landmass. Each day was allocated one (and only one) of these canonical patterns. From this canonical set, we selected $10$  prominent patterns, namely those that occurred at least once in four out of eight monsoon seasons from $2000-2007$ used as the reference data. Though we  defined such prominent patterns as those that appear at least once in four years, the actual frequency  was significantly higher. Likewise, the non-prominent patterns were much rarer than the cut-off employed.  These 10 patterns are enough to describe the spatial distribution of rainfall on about $95$ percent of the days.

In the current work, we analyze the spatial patterns in detail. Firstly, these patterns are robust, not only to parameter changes (see companion paper), but also they provide an accurate description of the monsoon, notably even for the period $1901-2000$, which is outside the reference data-set. Although each day is assigned only one of the prominent canonical patterns (i.e. patterns of Z state allocations), the actual discrete state allocation, at each individual location across India, for that day matches up very well (see Section \ref{subsec:robustness}). Moreover, mismatches between the assigned and observed rainfall patterns occur mostly in days of high rainfall (see Figure \ref{fig:hamm-sim}). Hence the spatial patterns identified may be used to \emph{define} typical rainfall behavior. Extreme high rainfall events are atypical in precisely the aforementioned sense.

The fact that these prominent canonical rainfall patterns reliably model the Indian summer monsoon seems to indicate that rainfall occurs in only a few characteristic miens across India (despite the monsoon being a complex multi-scale phenomenon). These patterns have ready interpretations as `scanty' or `abundant' rainfall ranging from $3-13\:\mbox{mm}$ per day over the entire Indian region. The dry patterns are seen to occur in higher frequency during the start and end of the monsoon period, though the occur throughout the monsoon. Note that a part of this effect may be explained by our choice to window the dataset based on the calendar date rather than dates of some meteorological import. The canonical spatial patterns can also be used to define all India `active' and `break' spells. In the literature, such spells have been identified using different approaches (mostly using fixed thresholds), yet our allocations of such spells broadly agree with their results.

In addition, we construct a low-dimensional Markov chain model by computing transition probabilities between prominent spatial patterns. To the best of our knowledge this is the first `ground-up' model of the spatio--temporal distribution of rainfall derived from data alone. This model can be used to simulate rainfall for a season reproducing the mean total rainfall as well as other characteristics. 

In the present work we also considered allocation of temporal patterns of rainfall to each location. These temporal patterns cluster spatial locations into regions with similar rainfall patterns in time. These patterns are then used to determine local active and break spells. We believe, regional definitions of active and break spells may be more relevant for agricultural and water-management purposes due to the highly heterogeneous nature of Indian monsoon.

One of the main goals of the current work has been to understand the spatial and temporal patterns of the discrete state allocation determined by our model. These patterns effectively cluster days and locations into groups with similar rainfall patterns. To benchmark our model, we also carried out the analysis using standard clustering algorithms such as K-means and spectral clustering. To the best of our knowledge, the present work is the first to consider clustering of Indian monsoon rainfall using these tools. Our work indicates that a fundamentally probabilistic approach is better suited to answer certain questions about spatio--temporal variability as well as distinguish underlying patterns of physical relevance. In particular we foresee the potential of the state transition matrix for spatial rainfall patterns as a predictive tool.

A question which naturally arises is why are these particular patterns in rainfall observed? In both the companion and current paper, we have adopted an exploratory outlook. To establish whether the patterns we discover have any meteorological significance necessarily takes us beyond the present analysis. However, the Markov Random Field model we employ may be readily extended to consider alternative physical variables such as outgoing long wave radiation (OLR), surface temperatures, fluid velocity, etc. Indeed it is even possible to construct a multi-variate version of the MRF model presented here thereby identifying combinations of patterns in different physical variables. Of course, such an analysis poses a significant computational challenge but is certainly feasible. Such an analysis will go beyond a simple understanding of correlations between meteorological variables but also provide greater statistical information such as, for example, transition probabilities. The advantage of the probabilistic approach presented is one can objectively define `typical' and `atypical' behavior. We seek to explain the rainfall patterns presented in the current work in terms of their connection to spatio-temporal patterns in other meteorological variables.  Such an analysis is the subject of ongoing work.

%%%%%%%%%%%%%%%%%%%%%%%%%%%%%%%%%%%%%%%%%%%%%%%%%%%%%
%%%%%%%%%%%%%%%%%%%%%%%%%%%%%%%%%%%%%%%%%%%%%%%%%%%%%
\section*{Acknowledgements}
AM was with ICTS-TIFR, Bangalore, India when most of this work was done. AM, AA, SV would like to acknowledge support of the Airbus Group Corporate Foundation Chair in Mathematics of Complex Systems established in ICTS-TIFR and TIFR-CAM. AM, AA would like to thank The Statistical and Applied Mathematical Sciences Institute (SAMSI), Durham, NC, USA where a part of the work was completed.

\bibliography{spatial-monsoon}
\bibliographystyle{plain}

%\printbibliography

\end{document}